\documentclass[letter]{aa} 

\usepackage{graphicx}

\usepackage{caption}
\usepackage{subcaption}

\usepackage{txfonts}
\begin{document} 



\title{UVIT Observations of the Star-Forming Ring in NGC7252: Evidence of Possible AGN Feedback Suppressing Central Star Formation}




\author{K. George\inst{1}\fnmsep\thanks{koshy@iiap.res.in},  P. Joseph\inst{1,2}, C. Mondal\inst{1}, A. Devaraj\inst{1}, A. Subramaniam\inst{1},  C. S. Stalin \inst{1}, P. C{\^o}t{\'e}\inst{3}, S. K. Ghosh\inst{4,5}, J. B. Hutchings \inst{3}, R. Mohan \inst{1}, J. Postma \inst{6}, K. Sankarasubramanian\inst{1,7}, P. Sreekumar \inst{1},  S.N. Tandon \inst{1,8}}


\institute{Indian Institute of Astrophysics, Koramangala II Block, Bangalore, India \and Department of Physics, Christ University, Bangalore, India \and National Research Council of Canada, Herzberg Astronomy and Astrophysics Research Centre, Victoria, Canada \and National Centre for Radio Astrophysics, Pune, India \and Tata Institute of Fundamental Research, Mumbai, India \and University of Calgary, Calgary, Alberta Canada \and  ISRO Satellite Centre, HAL Airport Road, Bangalore, India \and Inter-University Center for Astronomy and Astrophysics, Pune, India}


 
  \abstract
   {Some post-merger galaxies are known to undergo a starburst phase that quickly depletes the gas reservoir and turns it into a red-sequence galaxy, though the details are still unclear.} 
 {Here we explore the pattern of recent star formation in the central region of the post-merger galaxy NGC7252 using high resolution UV images from the UVIT on ASTROSAT. }
  {The UVIT images with 1.2 and 1.4 arcsec resolution in the FUV and NUV are used to construct a FUV-NUV colour map of the central region.}
   { The FUV–NUV pixel colour map for this canonical post-merger galaxy reveals a blue circumnuclear ring of diameter $\sim$ 10" (3.2 kpc) with bluer patches located over the ring. Based on a comparison to single stellar population models, we show that the ring is comprised of stellar populations with ages $\lesssim$  300 Myr, with embedded star-forming clumps of younger age ($\lesssim$ 150Myr).}
  {The suppressed star formation in the central region, along with the recent finding of a large amount of ionised gas, leads us to speculate that this ring may be connected to past feedback from a central super-massive black hole  that has ionised the hydrogen gas in the central $\sim$ 4" $\sim$ 1.3 kpc.}




\titlerunning{A Central Ring of Recent Star Formation in NGC7252}
\authorrunning{K. George\inst{1}}

\maketitle
%

\section{Introduction}

Elliptical galaxies in the local Universe are observed to be gas-poor systems with little or no detectable ongoing star formation. Since they are dominated by evolved stellar populations, they reside on the red sequence in the optical colour-magnitude relation \citep{Visvanathan_1977,Baldry_2004}. In the $\Lambda$CDM paradigm, the formation of such galaxies proceeds through a hierarchical process in which a spectrum of progenitors, having different gas contents and masses, merge during the protracted assembly of the galaxy \citep{Delucia_2006,Faber_2007,Naab_2009,Trujillo_2011}. \\

Generally speaking, major mergers are known to be common at high redshift while minor mergers dominate in the local Universe \citep{Conselice_2003,Kaviraj_2014}. According to the major merger scenario, two gas-rich star-forming disc galaxies can merge to form an elliptical galaxy \citep{Toomre_1972}. Initially, the merged system will be gas rich, but then undergoes a starburst phase that quickly depletes the gas reservoir and truncates further star formation.  In the high-redshift Universe, we observe many elliptical galaxies going through a starburst that follows a major merger. The manner in which star formation progresses in post-merger galaxies is, however, unclear: i.e., does it occur in a disc, or continue the star formation initiated from the pre-merger phase?


\begin{figure*}
\centering
\subcaptionbox{}{\includegraphics[width=0.252\textwidth]{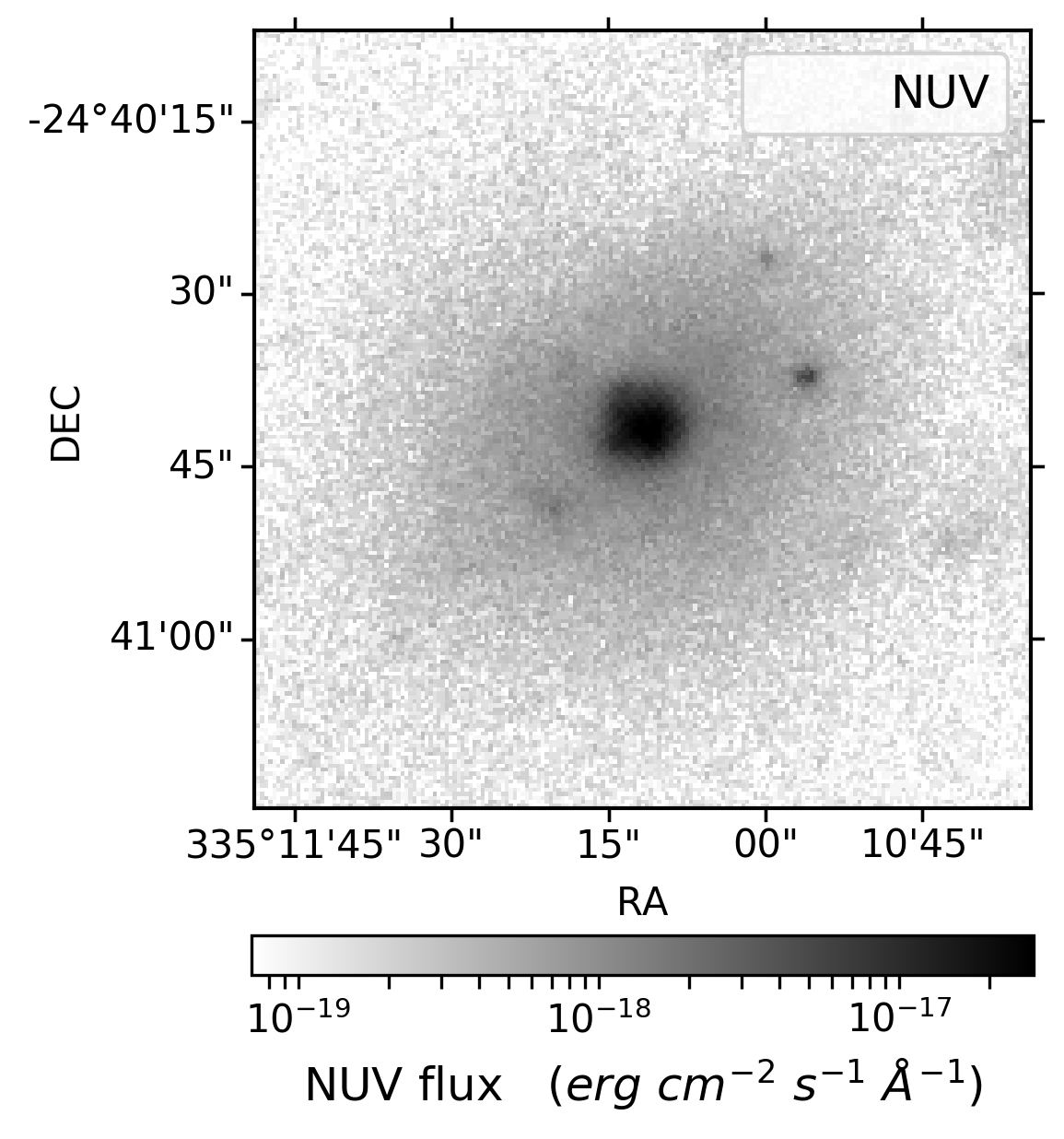}}%
\hfill 
\subcaptionbox{}{\includegraphics[width=0.252\textwidth]{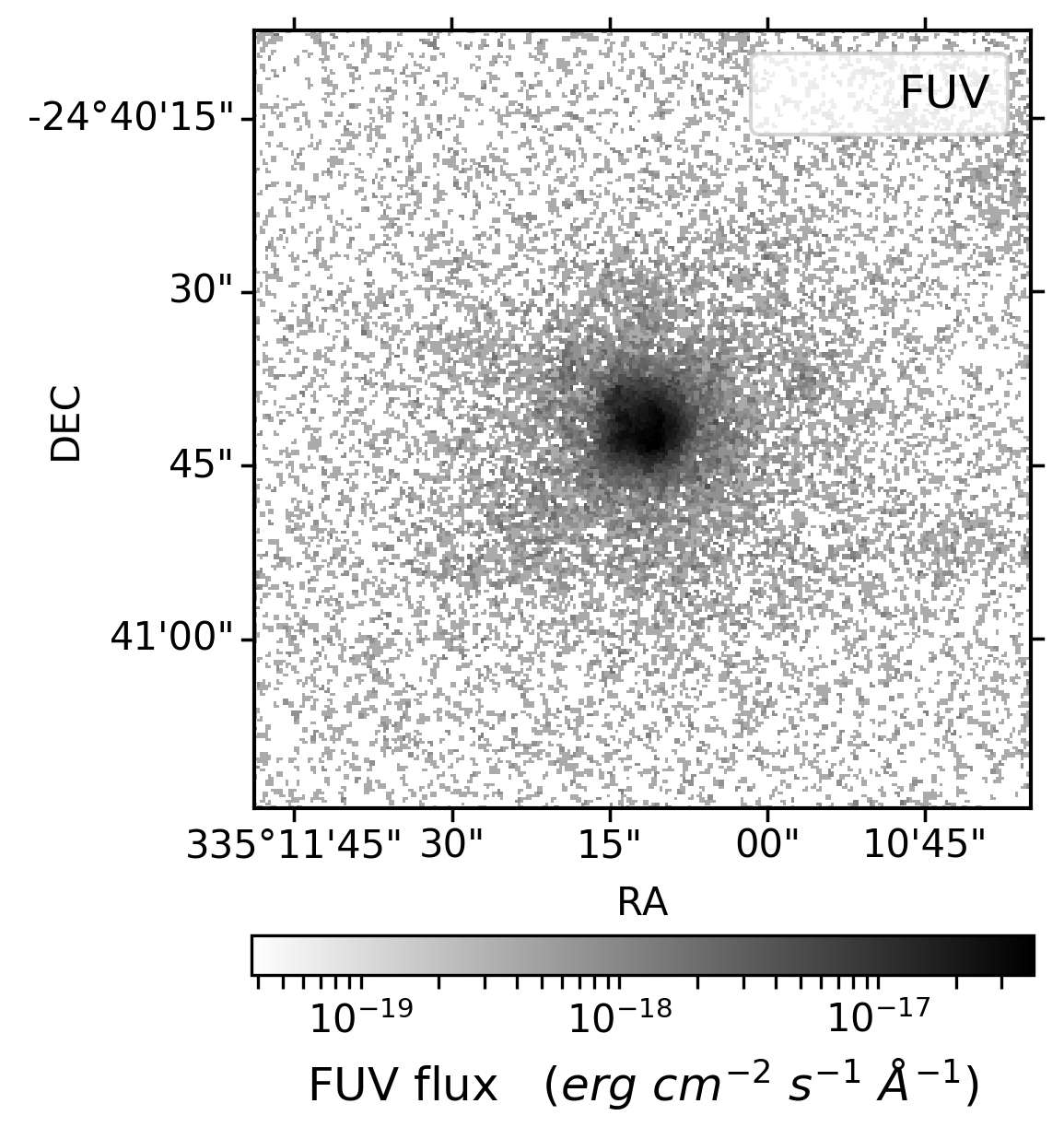}}%
\hfill 
\subcaptionbox{}{\includegraphics[width=0.209\textwidth]{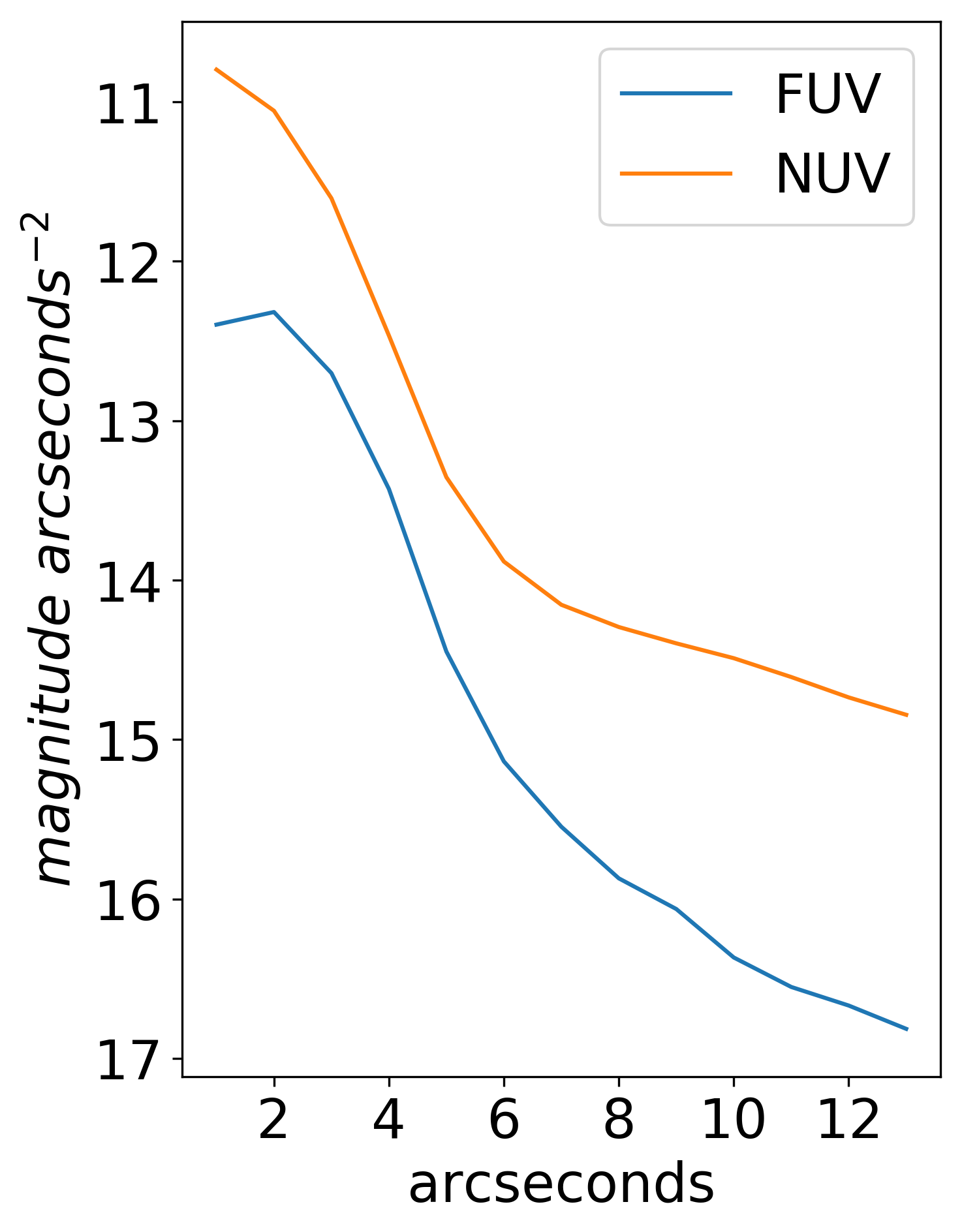}}%
\hfill 
\subcaptionbox{}{\includegraphics[width=0.261\textwidth]{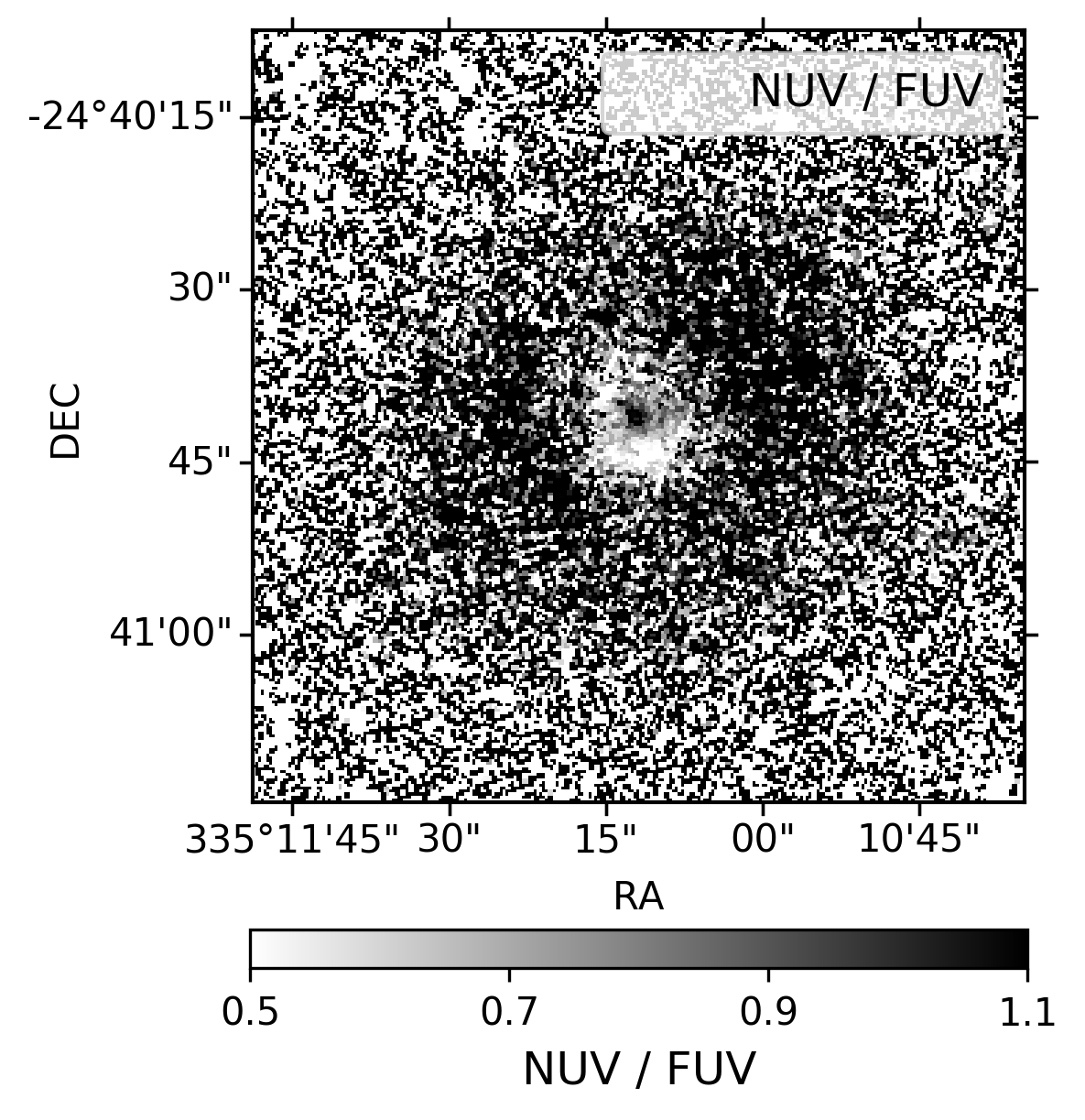}}%
\caption{The astrometry corrected NUV, FUV images (a and b), azimuthally averaged FUV, NUV profile (in concentric annuli of width 1") (c) and the NUV/FUV image of NGC7252 (d). The scaling of flux units is shown in the colour bar.}\label{figure:fig0}
\end{figure*}

There are  clues from observations of post-merger elliptical galaxies showing ongoing star formation; such systems often show signatures of disc components \citep{George_2017}. The post-merger galaxy NGC7252 is a classic example of an elliptical-like galaxy forming from the merger of two disc galaxies \citep{Schweizer_1982,Dupraz_1990,Wang_1992,Fritze_1994,Hibbard_1994}. The main body (i.e., central core) of NGC7252 has an elliptical morphology and exhibits intense star formation \citep{Schweizer_1982,Lake_1986,Hibbard_1995,Genzel_2001,Rothberg_2006}. The reservoir of gas acquired from the progenitor galaxies contributes to  ongoing star formation in a central disc, and early optical studies based on HST observations have demonstrated the existence of a small spiral pattern near the center of the galaxy {\citep{Dupraz_1990,Whitmore_1993,Miller_1997,Laine_2003,Rossa_2007,Ueda_2014} (also see Fig 1 of \citet{Weaver_2018}). 
NGC7252 is known to have a gaseous disc that is actively forming stars. The gas disc, which has a very low velocity dispersion (a condition necessary for star formation), formed within the last 100 Myr after final coalescence. It co-rotates with the stellar disc with an inward radial dependence on the ionisation state  \citep{Weaver_2018}. \\

We study the stellar disc in the main body of NGC7252 using newly acquired ultraviolet images. The integrated spectral energy distributions (SED) of young stellar populations peak at ultraviolet wavelengths due to the presence of hot OBA stars, so the UV flux directly traces the star formation within the past 100-200 Myr \citep{Kennicutt_2012}. The NUV and FUV images presented here  were acquired with the ultraviolet imaging telescope (UVIT) on board ASTROSAT, and thus probe recent star formation in this benchmark galaxy over scales of $\sim$400~pc.
Throughout this paper, we adopt a flat Universe cosmology with $H_{\rm{o}} = 71\,\mathrm{km\,s^{-1}\,Mpc^{-1}}$, $\Omega_{\rm{M}} = 0.27$, $\Omega_{\Lambda} = 0.73$ \citep {Komatsu_2011}.\\


\section{Observations and Analysis}


NGC7252\footnote{$\alpha$(J2000) = 22:20:44.7 and $\delta$(J2000) = 24:40:42 according to NED.} with a spectroscopic redshift $z \sim$ 0.0159 is located at a luminosity distance\footnote {http://www.astro.ucla.edu/wright/CosmoCalc.html} $\sim$ 68 Mpc \citep{Rothberg_2006}. The angular scale of 1" corresponds to 0.32 kpc at the distance of the galaxy. NGC7252 was observed at FUV (F148W filter, $\lambda_{mean}$=148.1nm, $\delta\lambda$=78.5nm) and NUV (N242W filter, $\lambda_{mean}$=241.8nm, $\delta\lambda$=50nm) wavelengths using the UVIT instrument on board the Indian multi-wavelength astronomy satellite ASTROSAT  \citep{Agrawal_2006}. The UVIT imaging yields a resolution of $\sim$ 1.2" for the NUV  and $\sim$ 1.4" for the FUV channels. Details on the UVIT observations of the star-forming tidal tail and main body of the galaxy are given in \citet{George_2018}. The astrometric calibration for the NUV and FUV images are performed using the {\tt astrometry.net} package where solutions are performed using the USNO-B catalogue \citep{Lang_2010}. The foreground extinction from the Milky Way Galaxy in the direction of NGC7252 is $A_{V}$ = 0.10 \citep{Schlegel_1998} which we scaled to the FUV and NUV $\lambda_{mean}$ values using the \citet{cardelli_1989} extinction law and corrected the magnitudes. The astrometry corrected NUV, FUV images, azimuthally averaged FUV, NUV profile and the ratio image of NGC7252 in flux units are shown in Figure~\ref{figure:fig0}. The NUV, FUV radial profiles in units of magnitude/arcsec$^{2}$ are created from concentric annuli of width 1", up to a distance of $\sim$ 13" from the center of the galaxy (AB magnitude system).  The NUV, FUV profiles and the ratio image  show higher values for NUV/FUV flux at the center of NGC7252. This implies a decrease in FUV flux towards the central region of NGC7252. We investigate this further in the following by creating a FUV-NUV colour map of the central region of NGC7252.\\

The NUV and FUV images of NGC7252 are first aligned in astrometry and then Gaussian-smoothed using a kernel of 1.4" in width to suppress noise and also to make the images match in resolution. The region of the FUV and NUV images that correspond to the main body of NGC7252 was isolated using the background counts from the whole image to set the threshold. Pixels with values above the 3$\sigma$ of the threshold were selected to isolate the region of interest. The counts in the selected pixels were background subtracted, integration time weighted, and converted to magnitude units using the zeropoints of \citet[][see their Table 4]{Tandon_2017b}. Magnitude for each pixel are used to compute the FUV--NUV colour map of the galaxy (see Figure~\ref{figure:fig1}). The pixels are colour coded in units of FUV--NUV colour. The image is of size $\sim$ 50$''$ $\times$ 50$''$ and corresponds to a physical size of $\sim$ 16~kpc on each side at the rest-frame of the galaxy. The corresponding azimuthally averaged FUV--NUV profile of the central region of the galaxy, which is shown in Figure~\ref{figure:fig2},  clearly shows a red core and surrounding blue ring. We note the following inferences; the FUV--NUV colour map of the main body of NGC7252 displays a red region at the center and outskirts of the galaxy, there exists a blue ring around the center with localised bluer patches.\\


We investigated if the ring is an artifact due to a mismatch in astrometry between the NUV and FUV images of NGC7252. We measured the WCS centroid positions of five field stars from the NUV and FUV images of the 28' UVIT field of view. The centroid offset between the images is found to be less than a pixel. We also created a pixel colour map of the five field stars which shows a smooth profile without any ring pattern. We thus confirm that the ring in the center region of NGC7252 is indeed significant.\\

The FUV--NUV colour map of NGC7252 main body reveals a circumnuclear ring that is significantly bluer than the central regions of the galaxy (i.e., FUV--NUV per pixel $\lesssim$ 0.5 mag). The ring is $\sim$ 3.2 kpc (10 arcsec) in diameter and contains bluer spots located on opposite sides. The central redder region is of size 1.3 kpc (4 arcsec), which is significantly redder in FUV--NUV colour as shown in Figure~\ref{figure:fig1}. The  FUV--NUV colour map of NGC7252 can be used to understand central star formation in this galaxy and can, in particular, offer insights into the last starburst. We used the {\tt Starburst99} stellar synthesis code to characterize the age of the underlying stellar population corresponding to the  FUV--NUV ring \citep{Leitherer_1999}. We selected 19 single stellar population (SSP) models over an age range of 1 to 900 Myr assuming a Kroupa IMF  \citep{Kroupa_2001} and solar metallicity (Z=0.02). The synthetic SED for a given age was then convolved with the effective area of the FUV and NUV passbands to compute the expected fluxes. The estimated values were then used to calculate the SSP ages corresponding to the observed FUV--NUV colours.  We performed a linear interpolation for the observed colour value and estimated the corresponding ages in all pixels in the FUV--NUV colour map. The age contours corresponding to 150, 250, 300 Myr were generated and overlaid over the colour map to isolate the regions of constant age, as shown in Figure~\ref{figure:fig1}. This exercise shows that the circumnuclear ring hosts stellar populations of age $\lesssim$  300 Myr, within which are embedded bluer, younger patches of star formation ($\lesssim$ 150-250 Myr). \\

Figure~\ref{figure:fig2} shows an azimuthally averaged colour profile in the central 12" of NGC7252. The FUV--NUV colour has been measured in concentric annuli of width 1" (320 pc). Note the striking change in the colour profile moving outwards, where the colour falls from red values in the core, to blue values in the ring, and finally to progressively redder colours with increasing distance from the galaxy center. \\

The FUV and NUV flux is subjected to extinction at the rest-frame of the galaxy. We do not have a proper extinction map of the galaxy NGC7252. The FUV--NUV pixel colour maps and the derived ages can be hence considered as the upper limits of the actual values.  We note that \citet[][see Figure 7a]{Weaver_2018} {\bf discusses} the extinction in the main body of the galaxy. \\


\section{Discussion}

The main body of NGC7252 is covered in the $\sim$ 40" region of the FUV--NUV colour map shown in Figure~\ref{figure:fig1}. The blue colour ring seen in FUV--NUV colour map is of size $\sim$ 10", similar to the one detected in previous studies. In the following, we call the blue patches seen on the ring as star-forming clumps. The ring can be caused either by high extinction or by inhibited star formation in the core (i.e central 4" of the galaxy). Because dust attenuates more flux in the FUV than in the NUV, FUV--NUV colours will be redder due to extinction. The A$_{V}$ map (in AB magnitude) created from $\mathrm{H}{\gamma}$/$\mathrm{H}{\beta}$ emission line ratio of the central $\sim$ 12" region, based on observations with IFU on VIMOS/VLT, is shown in Figure 7a of \citet{Weaver_2018}. The figure shows that the extinction at the core is small compared to the rest of the main body of the galaxy. But we note that the extinction is more than 2 mag in the eastern region between the young ($<$ 150 Myr) star forming clumps. It is therefore possible that the higher age ($<$ 250 Myr) of the star forming region between the young star-forming clumps can be due to the extinction. Hence the star forming clumps can be considered as the endpoints of star formation along an arc like structure. The Figure 7b of \citet{Weaver_2018} had demonstrated clearly that there is a lack of ongoing star formation in the core and the Figure 7a of \citet{Weaver_2018} shows that the extinction is negligible in the core. Taken together this rules out the possibility that the ring is caused by central extinction; rather, the central region seems devoid of recent star formation.\\

We independently confirm here the existence of a star forming ring with age $\lesssim$ 300 Myr, hosting clumps of younger star forming regions of age $<$ 150 Myr based on the UV imaging observations.  We also note that previous studies had demonstrated the existence of ionised and molecular gas rings \citep{Schweizer_1982,Wang_1992}. The previous observations on NGC7252 based on radio to X-ray wavelengths had so far not conclusively confirmed the presence of an AGN at the center. There exists [OIII] nebular emission 5 kpc south-west of the galaxy based on which \citet{Schweizer_2013} argue for the existence of a faint ionization echo excited by a mildly active nucleus that has declined by $\sim$ 3 orders of magnitude over the past 20,000–200,000 yr. The stellar and gaseous kinematics of the central 50" $\times$ 50" region of NGC7252 had been reported recently using the IFU on VIMOS/VLT \citep{Weaver_2018}. The study reports a central kpc-scale gas disc which has re-formed within the past 100 Myr since the formation of the central region of the galaxy. The present study confirms the presence of such a disc in the UV, but with the shape of a ring. The star forming ring is hosting multiple stellar populations of ages $<$ 150 Myr and $<$ 300 Myr, (even after taking into account the UV extinction between the young star forming clumps). This means that there were conditions responsible for triggering in-situ star formation at two epochs separated by $\sim$ 150 Myr. Also the ages of the star forming regions are compatible with the time required for the availability of cold gas in a settled disc near to the galaxy center after a major-merger \citep{Weaver_2018}. It is also interesting to note that the [OIII] emission line flux map of NGC7252 shows an enhancement in flux level at the central 2" region compared to the outskirts \citep{Weaver_2018}. The emission line flux maps cover the central 12" of the galaxy which incidentally covers the FUV--NUV blue colour ring we detect in NGC7252. The ionisation gas excitation is high inwards of 2" of the galaxy and is proposed to be due to shocks and AGN ionisation decreasing with distance from the center. This can be due to gas ioinisation from a low-luminosity AGN at the center. Also the drop in $\mathrm{H}{\beta}$ emission at the center of the galaxy confirms that the center is devoid of star formation.\\

The radio spectral index measured from the flux at 147 MHz and 1.4 GHz by TIFR GMRT Sky Survey (TGSS) and the NRAO VLA Sky Survey (NVSS) at the location of NGC7252 is found to be $\sim$ -0.66, typical of emission from AGN dominated by the extended lobes \citep{deGasperin_2018}. We note that the average peak-to-total TGSS flux ratio (0.80), the logarithm of TGSS flux density (-0.93 Jy) and the spectral index values from the direction of NGC7252 point towards the presence of AGN \citep[][see Figure 12]{deGasperin_2018}. We also note that previous X-ray observations had derived a photon index of 1.72 in the 0.5–10.0 keV from the nuclear region of NGC7252 \citep{Nolan_2004}. The derived X-ray photon index seems to suggest that the source is typical of a Seyfert type AGN \citep{Nandra_1994,Ishibashi_2010}. This can be considered as independent pieces of evidence coming from radio and X-ray wavebands about the nature of the source at the center of NGC7252.\\

\begin{figure}
\centering
\includegraphics[width=9.1cm,height=8.9cm,keepaspectratio]{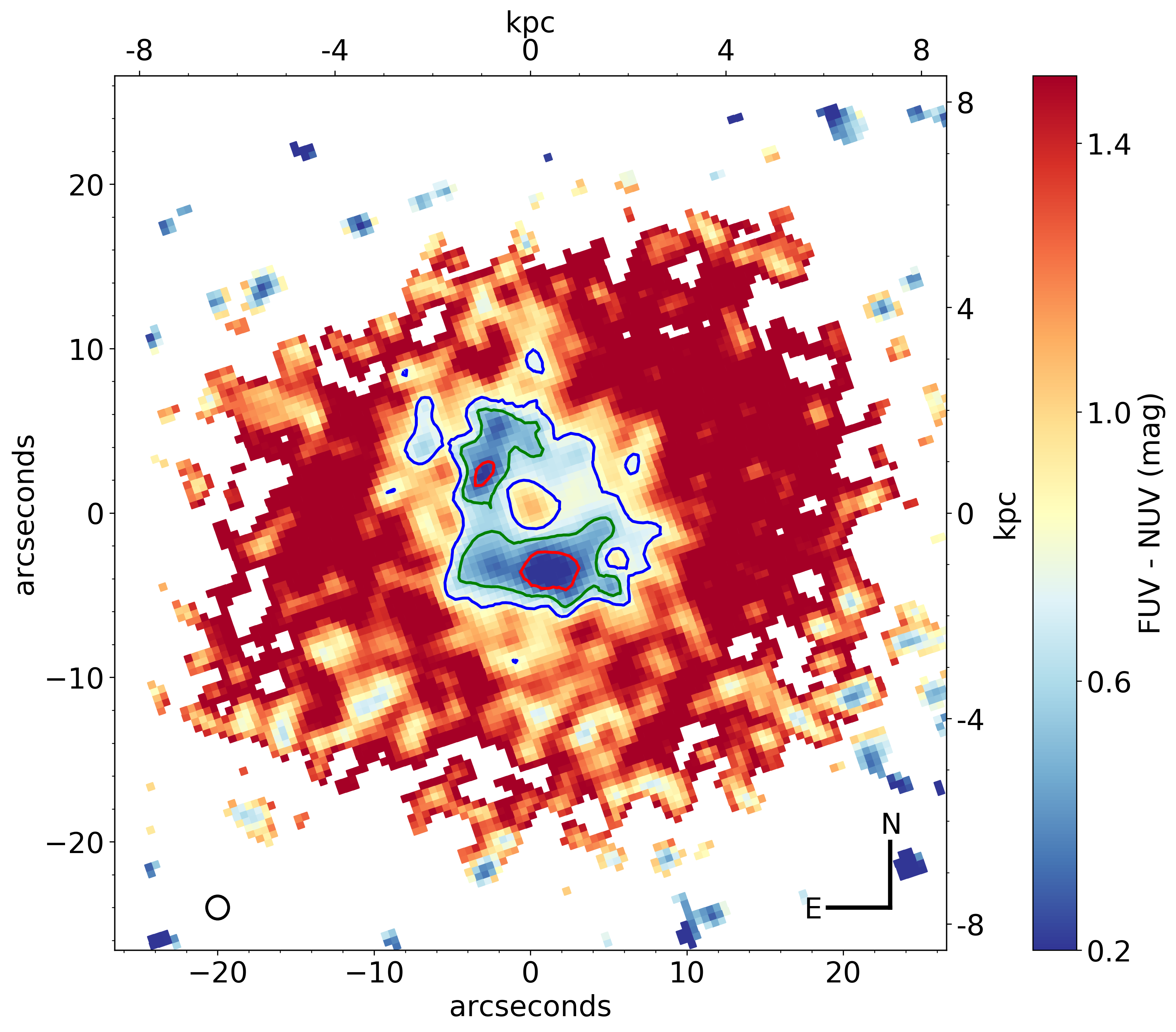}
\caption{$FUV-NUV$ colour map of the main body of NGC7252. The pixels are colour coded in units of $FUV-NUV$ colour. The point spread function for UVIT is shown in black circle. The image measures $\sim$ 50$''$ $\times$ 50$''$ and corresponds to a physical size of $\sim$ 16~kpc on each side. Age contours of 150 (red), 250 (green), 300 (blue) Myr are overlaid over the colour map to isolate regions of constant age. The blue ring is clearly seen with bluer colour clumps. The ring hosts young ($\lesssim$ 150 Myr) stellar populations compared to the rest of the galaxy.}\label{figure:fig1}
\end{figure}

\begin{figure}
\centering
\includegraphics[width=8.7cm,height=8.7cm,keepaspectratio]{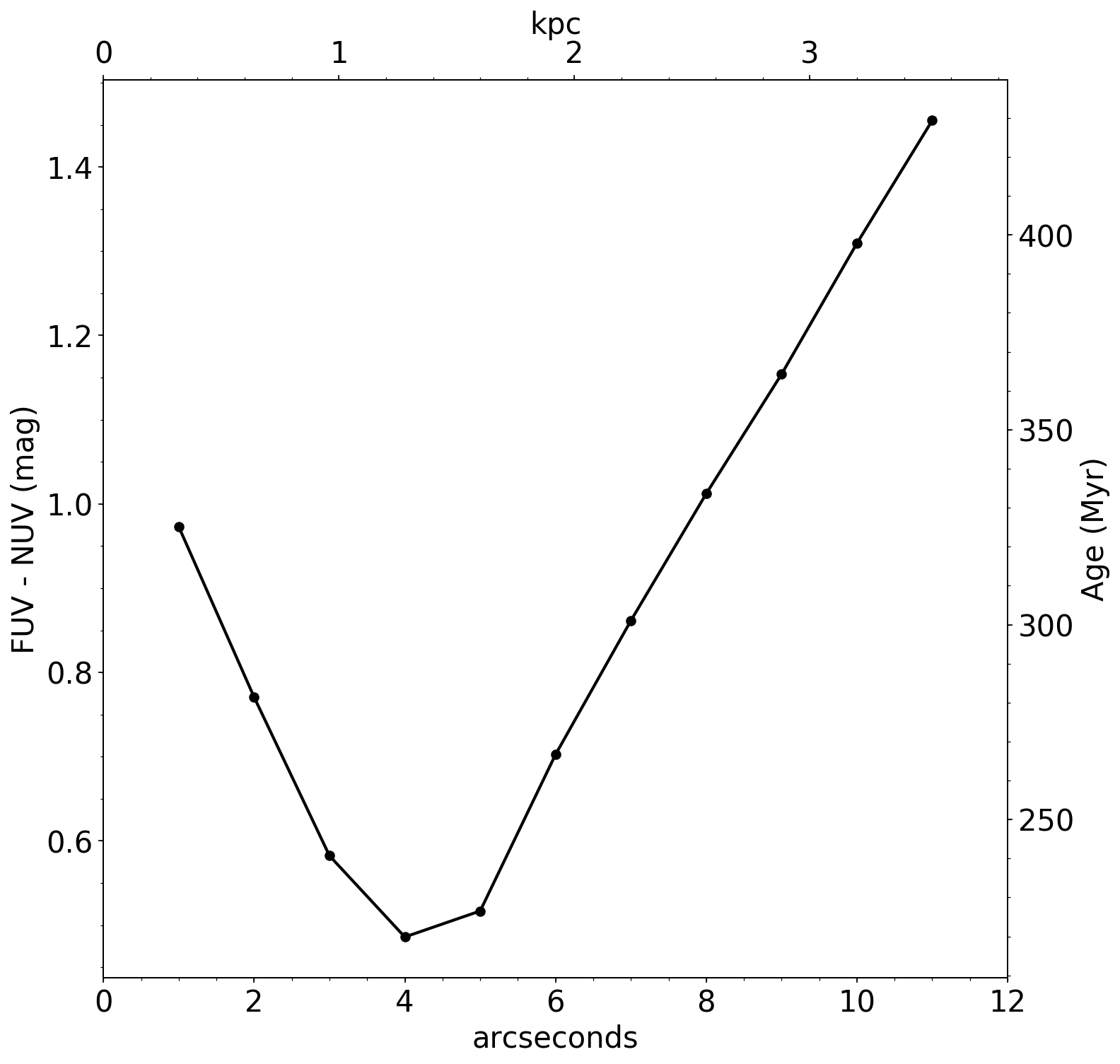}
\caption{Azimuthally averaged FUV--NUV colour profile of NGC7252. The 12" (3.8 kpc) region at the center of the galaxy been averaged in colour in concentric annuli of width 1". The profile shows negative gradient (i.e., an outward change from red to blue colours) out to $\sim$ 4", and then a steady reddening of the galaxy from 4" to 12". The FUV--NUV colour values and the corresponding age estimates are shown on the left and right axes, respectively.}\label{figure:fig2}
\end{figure}

The role of accreting black holes in suppressing star formation (quenching) and thereby regulating the galaxy evolution is studied extensively in recent literature using simulations and is also supported by observations \citep{Sanders_1988,Springel_2005,Dimatteo_2005,Schawinski_2007,Somerville_2008,Hopkins_2008,Schawinski_2010,Cheung_2016}. The starforming ring in NGC7252 is therefore interesting and could be carrying the imprint of star formation quenching in a post merger galaxy. The influence of a jet launched from an accreting black hole on the galaxy scale star formation is demonstrated in recent simulations \citep{Ishibashi_2012,Gaibler_2012,Ishibashi_2014}. The blast wave from the jet (launched horizontally) can create a bow shock in the vertical direction, pushing the gas outwards creating a cavity at the center. If the gas density reaches the critical density, triggered star formation can then proceed in a ring similar to the pattern we see in NGC7252 \citep[][see Figure 5]{Gaibler_2012}. We emphasise here that the ring-like star formation seen from the FUV--NUV colour map of NGC7252 can be reproduced in simulation where an accreting black hole can inhibit central star formation.\\

The FUV--NUV colour map reveals that star formation is not uniform along the ring, which is expected from a single burst of star formation. There exists young clumpy star forming regions with presumably strong dust-obscuration in between, which can be differentiated into a star forming arc with respect to the relatively older ring. This age-distinction can represent two epochs with conditions responsible for star formation. We propose two scenarios here; (a) due to the availability of cold gas following the merger at two separate periods, (b) the episodic activity of the central ionizing source inhibiting star formation (on the star forming ring) for a brief period. If true, similar episodic star formation activity can be expected which will delay the rapid depletion of the gas reservoir in NGC7252. This will further delay the (already higher) quenching time scale proposed for NGC7252  \citep{Weaver_2018} and slow down the migration of NGC7252 from the blue cloud to the red sequence on the galaxy colour-magnitude diagram.\\

The distribution of HI gas in NGC7252, based on VLA 21cm observations, has demonstrated that the neutral hydrogen gas is distributed more along the tidal tails compared to the center of the galaxy \citep{Hibbard_1994,Lelli_2015} (see Fig.3 of \citet{Lelli_2015} for HI contours overlaid over the $R-$band image of NGC7252.). The gas at the center might have been acquired in the merger, settling in the form of a disc that hosts star formation and perhaps feeds a possible central black hole.  The blue circumnuclear ring seen in the  FUV--NUV colour map may then be the signature of the last burst of star formation fed by gas left over from the merger. The main body of this post-merger galaxy shows properties typical of elliptical galaxies, but resides in the blue cloud of the galaxy colour-magnitude diagram \citep{Weaver_2018}. We note that if NGC7252 were at higher redshifts the tidal tails would be very faint to detect in imaging data and the galaxy would be classified as a star forming blue early-type galaxy \citep{Schawinski_2009,George_2017}. In this scenario, it is expected that NGC7252 will evolve onto the red sequence once the ongoing star formation is quenched. \\






\section{Conclusions}

The post-merger galaxy NGC7252 has been observed with the UVIT instrument on ASTROSAT. FUV and NUV images with a spatial resolution of 1.2"--1.4" have been used to construct the first FUV--NUV colour map of the galaxy. We detect a blue, circumnuclear ring of diameter $\sim$ 10" (3.2 kpc). The central 4" (1.3 kpc) region of the galaxy has a red FUV--NUV colour compared to the blue ring. We used single stellar population models, converting the FUV--NUV pixel colour map into an age map, to demonstrate that this star-forming ring hosts stellar populations age $\lesssim$ 300 Myr, with bluer clumps of younger star-forming regions ($\lesssim$ 150 and 250 Myr) distributed over the ring. Results from previous studies suggest the existence of an ionising source at the center of the galaxy, likely an accreting black hole that was active in the past. The presence of such a source in the center of this post-merger galaxy would have a negative feedback on surrounding star formation (which requires dynamically cold molecular hydrogen). Star formation in the core would thus be inhibited, while triggered star formation proceeds in a surrounding ring.





\begin{acknowledgements}
This publication uses the data from the AstroSat mission of the Indian Space Research  Organisation  (ISRO),  archived  at  the  Indian  Space  Science  Data Centre (ISSDC).  
This research made use of Astropy, a community-developed core Python package for Astronomy \citep{Astropy_Collaboration_2018}.
\end{acknowledgements}


\end{document}